\documentstyle[a4wide,12pt]{article}
\begin{document}
\begin{flushright}
\baselineskip=12pt
{SUSX-TH-98-006}\\
{hep-th/yymmnn}\\
{March 1998}
\end{flushright}
\begin{center}
{\LARGE \bf CP-violating phases in the CKM matrix in orbifold
compactifications}
\vglue 0.45 cm

{D. BAILIN$^{\clubsuit}$ \footnote
{D.Bailin@sussex.ac.uk}, G.V. KRANIOTIS$^{\spadesuit}$ \footnote
{G.Kraniotis@rhbnc.ac.uk} and A. LOVE$^{\spadesuit}$ \\}
{$\clubsuit$ \it Centre for Theoretical Physics, \\}
{\it University of Sussex,\\}
{\it Brighton BN1 9QJ, U.K. \\}
{$\spadesuit$ \it Department of Physics, \\} 
{\it Royal Holloway and Bedford New College. \\}
{\it University of London, Egham, \\}
{\it Surrey TW20-0EX, U.K. \\}
\baselineskip=12pt

\vglue 0.25cm
ABSTRACT
\end{center}
{\rightskip=3pc
\leftskip=3pc
\noindent
\baselineskip=20pt

The picture of $CP$-violation in orbifold compactifications 
in which the $T$-modulus is at a complex fixed point of the modular group
is studied.  $CP$-violation  in the neutral kaon system and in the neutron 
electric dipole moment are both discussed. The situation where the
$T$-modulus 
 takes complex values on the unit circle which are not 
at a fixed point is also
discussed.

\newpage

 String theory may provide new insights into the origin of $CP$-violation. It 
has been argued \cite{Dine} that there is no explicit $CP$ symmetry breaking 
in string theory either at the perturbative or the non-perturbative level. 
However, spontaneous $CP$-violation may arise from complex expectation values 
of moduli or other scalars \cite{Dine, Iblust, Brig, Achar, BKL1, BKL2}. In
any 
supergravity theory there is the possibility of $CP$-violating phases in
the soft
 supersymmetry-breaking $A$ and $B$ terms and in the gaugino masses, which
are 
in addition to a possible phase in the CKM matrix and the $\theta$
parameter of QCD.
(For a review with extensive references to the earlier literature see
reference 
\cite{Gross}.)

In compactifications of string theory, soft supersymmetry-breaking terms
can be
 functions of moduli such as those associated with the radii and angles 
characterising the underlying torus of an orbifold compactification.
Moreover, 
in general, the Yukawa couplings also depend upon these moduli. Thus, if the 
orbifold moduli develop complex expectation values, this can feed through
 into the corresponding low energy supergravity as potentially $CP$-violating 
phases in the soft supersymmetry-breaking terms and the Yukawa couplings, 
and, after transforming to the quark mass eigenstates, in the CKM matrix.

The $CP$-violating phases in the soft supersymmetry-breaking terms lead 
to a non-zero neutron electric dipole moment, and, if these phases are larger
 than about $10^{-3} - 10^{-2}$, this electric dipole moment will be above the
 experimental upper bound. They also contribute to the $CP$-violating
parameter
 $\epsilon_K$ in the neutral kaon system through the squark mass matrix, and
phases of order $1$ in the soft supersymmetry-breaking terms can easily
violate 
bounds derived from the experimental value of $\epsilon_K$ by as much 
as seven orders of magnitude \cite{Gross}. It is therefore important to check 
that any phases in the soft supersymmetry-breaking terms arising from complex
 values of the orbifold moduli are not so large as to produce too large a
neutron
 electric dipole moment or too large an $\epsilon_K$ parameter in the neutral 
kaon system.

On the other hand, it is an attractive possibility that complex orbifold
moduli
 may induce phases in the Yukawa couplings that emerge as $CP$-violating
phases 
in the CKM matrix. Our knowledge of the weak mixing angles suggests that we
require
 the largest phase in the CKM matrix to be of order $10^{-1} - 10^{0}$ with
the 
relatively small amount of $CP$-violation in the neutral kaon system being
due to
small mixing angles. It is a challenge for any model of $CP$-violation to
square 
the large phase required in the CKM matrix, and so in the Yukawas,
 with the small phases required in the soft supersymmetry-breaking terms,
 as discussed above. (It might, however, be possible to have only small
$CP$-violating
phases in the CKM matrix if the soft supersymmetry-breaking terms are
non-universal
and $\epsilon_K$ is due to squark-gluino box diagrams \cite{Gross}.)

Elsewhere \cite{Achar, BKL1, BKL2}, we studied the expectation values 
of orbifold $T$-moduli by minimising the modular invariant effective
potential.
In these calculations, the greatest uncertainty relates to the dilaton
dynamics,
 since, as is well known, a single gaugino condensate superpotential does not 
stabilize the dilaton expectation value at a realistic value. One approach we 
adopted, following earlier authors \cite{FMTV}, was to simulate the dilaton 
dynamics by treating the dilaton expectation value $S$ and its corresponding 
auxiliary field $F_S$ as free parameters, but with $ReS$ fixed at 2. In
line with 
existing calculations of $S$ and its auxiliary field in various contexts,
we assumed
that $S$ and $F_S$ were both real. Another approach we adopted was to assume 
stabilisation of the dilaton expectation value by a general multiple
gaugino condensate
or by a non-perturbative dilaton K${\rm {\ddot a}}$hler potential.
 The outcome, in any of these approaches, for the case of a single 
overall modulus $T$, is that there was some region of parameter space
 for which the minimum of the effective potential gave a real value 
of $T$ different from 1. There were other regions of parameter space for which
the minimum was either at the $PSL(2,Z)$ modular group fixed point at $T=1$
 or at the fixed point at $T=e^{i\pi/6}$.

When the non-perturbative superpotential was generalised to contain the
absolute modular
 invariant function $j(T)$ as well as the Dedekind eta function $\eta(T)$
there 
were other possibilities in the case of a general multiple gaugino
condensate. For 
some choices of the parameter
\begin{equation}
\rho=\frac{1-F_S}{y}
\label{ro}
\end{equation}
where
\begin{equation}
y=S+\bar{S}-{\tilde{\delta}}_{GS}\log(T+\bar{T})
\label{dil}
\end{equation}
with ${\tilde{\delta}}_{GS}$ the Green-Schwarz coefficient, there were also
minima
 where $T$ was on the boundary of the fundamental domain at a point different
 from $1$ or $e^{i\pi/6}$. Values of $T$ inside the fundamental domain
could also
 be obtained in the case of a non-perturbative dilaton K${\rm {\ddot
a}}$hler potential
 , but only for choices of the K${\rm {\ddot a}}$hler potential that gave
the wrong 
sign kinetic terms.

The implications for $CP$-violating phases in the soft
supersymmetry-breaking terms
 can be seen by studying the following expressions in terms of the overall
modulus $T$.
 Allowing for the possibility of stringy non-perturbative corrections
\cite{Casas} to the 
dilaton K${\rm {\ddot a}}$hler potential, we write the dilaton and
moduli-dependent 
part of the  K${\rm {\ddot a}}$hler potential as
\begin{equation}
K=P(y)-3\log(T+\bar{T})
\end{equation}
where $y$ is given by (\ref{dil}). Then $\frac{dP}{dy}$ and $\frac{d^2 P}{d
y^2}$, 
which occur in the soft supersymmetry-breaking terms, are treated as
parameters.
 Also, allowing for a general multiple gaugino condensate, we write the
 non-perturbative superpotential in the form 
\begin{equation}
W_{np}=\Omega(\Sigma)[\eta(T)]^{-6}
\end{equation}
where 
\begin{equation}
\Sigma=S+2\tilde{\delta}_{GS} \log \eta(T)
\end{equation}
and
\begin{equation}
\Omega(\Sigma)=\Sigma_a h_a e^{\frac{24\pi^2}{b_a} \Sigma}
\end{equation}
Then $\Sigma$ is a parameter to be chosen so that $y$ is approximately $4$. 
The parameter $\rho$ of (\ref{ro})is related to $\Omega$ by
\begin{equation}
\rho = \frac{1}{\Omega} \frac{d\Omega}{d\Sigma}
\end{equation}
Then the gaugino masses $M_a$ are
\begin{eqnarray}
M_a&=&m_{3/2}(Re f_a)^{-1}\Bigl[\frac{\partial \bar{f_a}}{\partial \bar{S}}
(\frac{d^2 P}{d y^2})^{-1}( \frac{dP}{dy}+\rho) \nonumber \\ 
&+ &(\frac{b^{'}_a}{8\pi^2} -\tilde{\delta}_{GS}
)(1+\frac{1}{3}\tilde{\delta}_{GS} \frac{dP}{dy})^{-1}
(\rho \frac{1}{3}\tilde{\delta}_{GS} -1)(T+\bar{T})^2 |\hat{G}|^2 \Bigr]
\nonumber \\  
\end{eqnarray}
where 
\begin{equation}
\hat{G}(T,\bar{T})=(T+\bar{T})^{-1}+2\eta ^{-1} \frac{d\eta}{dT}
\end{equation}
and $b'_a$ is the usual coefficient occurring in the string loop threshold
corrections
 to the gauge coupling constants \cite{iblust92,DKL, DFKZ}. Provided the
dilaton
 auxiliary field $F_S$ is real, as we have discussed earlier, there are no 
$CP$-violating phases in the gaugino masses.

The soft supersymmetry-breaking $A$ terms are given by
\begin{eqnarray}
m_{3/2}^{-1} A_{\alpha \beta \gamma} &=& (\frac{d^2 P}{d y^2})^{-1}
(\frac{dP}{dy}+\rho) \frac{dP}{dy} \nonumber \\ 
&+&\left(1+\frac{1}{3}\tilde{\delta}_{GS} \frac{dP}{dy} \right)^{-1}(1-\rho
\frac{\tilde{\delta}_{GS}}{3})
(T+\bar{T})\bar{\hat{G}}  \nonumber \\
&\times& (3+n_{\alpha}+n_{\beta}+n_{\gamma} - (T+\bar{T})\frac{\partial \log 
h_{\alpha \beta \gamma}}{\partial T}) \nonumber \\ 
\label{aterm}
\end{eqnarray}
where the superpotential term for the Yukawa couplings of $\phi_{\alpha},
\phi_{\beta}$ and 
$\phi_{\gamma}$ is $h_{\alpha \beta
\gamma}\phi_{\alpha}\phi_{\beta}\phi_{\gamma}$, the 
modular weights of the states are  $n_{\alpha}, n_{\beta}$ and
$n_{\gamma}$, and the usual 
rescaling by a factor  $\frac{|W_{np}|}{W_{np}} e^{K/2}$ has been carried
out to go from the
 supergravity theory derived from the orbifold compactification to the
spontaneously
 broken globally supersymmetric theory \cite{Brig}.

The soft supersymmetry-breaking $B$ term depends upon the mechanism assumed
for generating
 the $\mu$ term of the Higgs scalars $H_1$ and $H_2$, with corresponding
superfields 
$\phi_1$ and $\phi_2$. If we assume that the $\mu$ term is generated
non-perturbatively as 
an explicit superpotential term $\mu_W \phi_1 \phi_2$ then the $B$ term,
which in this case 
we denote by $B_W$, is given by 
\begin{eqnarray}
m_{3/2}^{-1} B_W &=& -1+(\frac{d^2 P}{d y^2})^{-1}
(\frac{dP}{dy}+\bar{\rho})(\frac{dP}{dy}+\frac{\partial \log
\mu_W}{\partial S}) \nonumber \\ 
&+&(1+\frac{1}{3}\tilde{\delta}_{GS} \frac{dP}{dy})^{-1}(1-\bar{\rho} 
\frac{\tilde{\delta}_{GS}}{3})
(T+\bar{T})\bar{\hat{G}}  \nonumber \\
&\times& \left(3+n_1+n_2 - (T+\bar{T})\frac{\partial \log \mu_W}{\partial T}-
\tilde{\delta}_{GS}\frac{\partial \log \mu_W}{\partial S}\right)
 \nonumber \\ 
\label{bwterm}
\end{eqnarray}
where $n_1$ and $n_2$ are the modular weights of the Higgs scalar
superfields, and
again the appropriate rescaling has been carried out.

On the other hand, if the $\mu$ term is generated by a term of the form
$Z\phi_1 \phi_2 +h.c.$
 , with $Z$ real, mixing the Higgs superfields in the in the K${\rm {\ddot
a}}$hler potential
 \cite{AGNT}, then (before rescaling the Lagrangian) the $B$ term, which in
this case
 we denote by $B_Z$, is given by
\begin{eqnarray}
m_{3/2}^{-1}\mu^{eff}_Z B_Z &=& W_{np}
Z\Bigl[2+\Bigl((T+\bar{T})(3+\tilde{\delta}_{GS} \frac{dP}{dy})^{-1}
(\rho \tilde{\delta}_{GS}-3)\hat{G}(T,\bar{T})+h.c.\Bigr)\Bigr] \nonumber \\
&+& W_{np}Z \Bigl[-3+|\frac{dP}{dy}+\rho|^2 (\frac{d^2 P}{d y^2})^{-1}
\nonumber \\
&+& (3+\tilde{\delta}_{GS} \frac{dP}{dy})^{-1} |\rho \tilde{\delta}_{GS}-3|^2
(T+\bar{T})^2 |\hat{G}(T,\bar{T})|^2\Bigr]
\label{bzterm}
\end{eqnarray}
The effectve $\mu$ term in the superpotential is $\mu=\mu^{eff}_Z$ where
\begin{equation}
 \mu^{eff}_Z= e^{K/2}|W_{np}|Z(1-\frac{1}{3}\rho \tilde{\delta}_{GS})
\left(1-(T+\bar{T})\hat{G}(T,\bar{T})\right)
\label{muf}
\end{equation}
To obtain the final form for $B_Z$ in the low energy supersymmetry theory,
rescaling
 of the Lagrangian by $\frac{|W_{np}|}{W_{np}} e^{K/2}$ has to be carried out.

Finally, the soft supersymmetry-breaking scalar masses squared, which are
always real, 
are given by 
\begin{equation}
m^2_{\alpha}=V_0+m^2_{3/2}+m^2_{3/2}n_{\alpha}
\left(1+\frac{1}{3}\tilde{\delta}_{GS} \frac{dP}{dy} \right)^{-2}
|1-\frac{1}{3}\tilde{\delta}_{GS} \rho|^2 (T+\bar{T})^2
\left|\hat{G}(T,\bar{T})\right|^2
\label{scalar}
\end{equation}
where $V_0$ is the vacuum energy.

Whenever the value of $T$ at the minimum of the effective potential is at a
zero
of $\hat{G}(T,\bar{T})$, all $CP$-violating phases in the soft
supersymmetry-breaking
 terms are zero, provided $F_S$ is real, so that the parameter $\rho$ is
also real.
 Moreover, the $A$ terms and the soft supersymmetry-breaking scalar masses
are universal.
 This is evident from (\ref{aterm}) and (\ref{scalar}) because the
dependence on the 
modular weights $n_{\alpha}$ and the Yukawas $h_{\alpha \beta \gamma}$
drops out of the 
expressions. In addition, the gaugino masses are universal at the
unification scale for 
the coupling constants because $Ref_a = g_a^{-2}$ and because the only $S$
dependence
 of $f_a$ is in the tree term $f_a=S$.

The function $\hat{G}(T,\bar{T})$ has zeros at the fixed points 
$T=1$ and $T=e^{i\pi/6}$ of the $PSL(2,Z)$ modular group. The fixed point at
 $T=e^{i\pi/6}$ is particularly interesting. Despite the large phase of $T$,
 the phases $\phi(B)$ and $\phi(A)$ of the $B$ term and the universal
$A$ term are zero. In these circumstances, there is no contribution to 
the electric dipole moment of the neutron \cite{Gross} from the complex
 expectation value of the $T$ modulus. Moreover, because the soft
supersymmetry-breaking
 terms are universal, there is no contribution to $\epsilon_K$ from the
squark mass matrix
\cite{Gross}. Any minimum obtained from $T=e^{i\pi/6}$ by a modular
transformation is 
also a zero of $\hat{G}(T,\bar{T})$, and the same discussion applies.

The next question that needs to be considered is whether the phase of the
$T$ modulus
 can induce the phase required in the CKM matrix to account for the size of
the
$CP$ violation in the kaon system. In general, the quark mass matrix is given
 by the Lagrangian terms
\begin{equation}
{\cal L}_M=(h_d)_{fg}(\bar{d}_f)_L(d_g)_Rv_1-(h_u)_{fg}(\bar{u}_f)_L(u_g)_Rv_2
\end{equation}
where $f,g=1,2,3$ label the quark generations and 
$v_1,v_2$ are Higgs expectation values. 
If the transformation to the quark mass eigenstates is
\begin{equation}
d_L=P_L\tilde{d}_L, u_L=Q_L\tilde{u}_L, d_R=P_R\tilde{d}_R, u_R=Q_R\tilde{u}_R
\end{equation}
then the charged weak current in terms of the quark mass eigenstates is
\begin{equation}
J^{\mu}_{+}=\bar{\tilde{u}}_L\gamma^{\mu}V\tilde{d}_L
\end{equation}
where 
\begin{equation}
V=Q^{\dagger}_LP_L
\label{ckm}
\end{equation}
is the CKM matrix. In terms of the real diagonal matrices $\tilde{M}_u$ and 
$\tilde{M}_d$ of physical quark masses
\begin{equation}
V=-v_1v_2(\tilde{M}_u)^{-1}Q_R^{\dagger}h_u^{\dagger}h_dP_R(\tilde{M}_d)^{-1}
\end{equation}
where the Higgs expectation values have been chosen real. Generically, the
largest
 phase in the CKM matrix $V$ is of the same order as the largest phase in
the Yukawa
 matrices $h_u$ and $h_d$. (An explicit example of the way in which the
phases in the 
Yukawas feed through into the CKM matrix can be found in \cite{CM}, where
an orbifold-
compactification-inspired parameterisation of the Yukawa matrix containing
6 phases is
 employed.)

To study the single overall modulus case, we restrict attention to the $Z_M
\times Z_N$
orbifolds, which are the only orbifolds \cite{DKL} which possess three
$N=2$ moduli
 $T_i, i=1,2,3$. For these orbifolds, it is known that the contributions
$h_i$ to the
 (unnormalised) Yukawa couplings $h$ from the modulus $T_i$ take one of
five forms
 \cite{BLS}. (We are assuming that the phases in the CKM matrix derive from
renormalisable
 couplings.) First, $h_i$ may have no moduli dependence if either one or
more of the 
three coupled states is untwisted, or the 3-point function reduces to a
2-point function 
because the $i$th complex plane is unrotated in one of the three twisted
sectors involved.
 For all other cases we write
\begin{equation}
h_i\sim\sum_{X^i_{cl}}exp(-S^i_{cl})
\end{equation}
where $S^i_{cl}$ is the classical action continued to Euclidean space, and
\begin{equation}
h=\prod _{i=1}^{3}h_i
\end{equation}
For the $Z_3 \times Z_3$ orbifold, when all 3 twists in the $i$th complex
plane are
 $e^{2\pi i/3}$, then
\begin{equation}
h_i\sim\sum_{k_{2i-1},k_{2i}}
\exp\Bigl(-\frac{1}{6}\pi T_i g_i (k_{2i-1},k_{2i})\Bigr)
\label{yuk1}
\end{equation}
where
\begin{equation}
g_i(k_{2i-1},k_{2i})=(2p_{2i-1}+3k_{2i-1}+6k_{2i})^2+3k_{2i-1}^2
\end{equation}
$k_{2i-1},k_{2i}$ are arbitrary integers, and $p_{2i-1}=0,\pm1,\pm2$
depending on the fixed tori involved.

For the $Z_2 \times Z_6, Z_3 \times Z_6, Z_2 \times Z_6'$ and $Z_6 \times
Z_6$ orbifolds,
 in addition to (\ref{yuk1}), two other forms of Yukawa contribution arise. 
For one twist of $e^{4\pi i/3}$ and two twists of $e^{2\pi i/6}$
\begin{equation}
h_i\sim\sum_{k_{2i-1},k_{2i}}
\exp \left(-\frac{1}{12}\pi T_i g_i (k_{2i-1},k_{2i})\right)
\label{yuk2}
\end{equation}
with $p_{2i-1}=0,\pm1$. For twists of $e^{2\pi i/6}, e^{2\pi i/3}$ and $-1$
\begin{equation}
h_i\sim\sum_{k_{2i-1},k_{2i}}
\exp\left(-\frac{1}{12}\pi T_i e_i (k_{2i-1},k_{2i})\right)
\label{yuk3}
\end{equation}
where 
\begin{equation}
e_i(k_{2i-1},k_{2i})=(q_i+6k_{2i})^2+3(t_i+6k_{2i}+4k_{2i-1})^2
\end{equation}
with
\begin{eqnarray}
q_i&=&2n_{2i-1}+2p_{2i} \nonumber \\
t_i&=&2n_{2i-1}+3p_{2i}-2p_{2i-1}
\end{eqnarray}
and the integers $n_{2i-1}$ taking the values $0,\pm1$, and the integers
$p_{2i}$ and 
$p_{2i-1}$ taking the values $0,1$, depending on the fixed tori involved.
Finally, for the
$Z_2 \times Z_4$ and $Z_4 \times Z_4$ orbifolds, the only moduli dependent
contributions 
to the Yukawa couplings occur for two twists of $e^{2\pi i/4}$ and one
twist of $-1$. Then
\begin{equation}
h_i\sim\sum_{k_{2i-1},k_{2i}}
\exp\left(-\frac{1}{4}\pi T_i l_i (k_{2i-1},k_{2i})\right)
\label{yuk4}
\end{equation} 
where 
\begin{equation}
l_i(k_{2i-1},k_{2i})=(\tilde{q}_i+4k_{2i-1}+2k_{2i})^2+(\tilde{t}_i-2k_{2i
})^2
\end{equation}
with
\begin{eqnarray}
\tilde{q}_i&=&2\tilde{n}_{2i}+2\tilde{p}_{2i-1}-\tilde{p}_{2i} \nonumber\\
\tilde{t}_i&=&\tilde{p}_{2i}
\end{eqnarray}
and the integers $\tilde{n}_{2i},\tilde{p}_{2i-1}$ and $\tilde{p}_{2i}$ all
taking 
the values $0,1$, depending on the fixed tori involved.

The above contributions to the Yukawas may be cast in terms of Jacobi
$\theta$ 
functions $\theta_3$ and $\theta_2$ where
\begin{eqnarray}
\theta_3(\nu,i \gamma T)&=&\sum_{n=-\infty}^{\infty}e^{-\pi \gamma Tn^2}e^{2\pi
in\nu} \nonumber \\
\theta_2(\nu,i \gamma T)&=&\sum_{n=-\infty}^{\infty}e^{-\pi \gamma
T(n-1/2)^2}e^{2\pi i(n-1/2)\nu} \\
\end{eqnarray}
Then, corresponding to equations (\ref{yuk1}),(\ref{yuk2}),(\ref{yuk3}) and
(\ref{yuk4}), 
respectively we have

\begin{eqnarray}
h_i&\sim& e^{-2\pi p_{2i-1}^2
T_i/3}\Bigl[\theta_3(ip_{2i-1}T_i,2iT_i)\theta_3(ip_{2i-1}T_i,6iT_i)\nonumber\\
&+&\theta_2(ip_{2i-1}T_i,2iT_i)\theta_2(ip_{2i-1}T_i,6iT_i)\Bigr] \nonumber\\
\label{yuka1}
\end{eqnarray}
\begin{eqnarray}
h_i&\sim& e^{-\pi p_{2i-1}^2
T_i/3}\Bigl[\theta_3(ip_{2i-1}T_i/2,iT_i)
\theta_3(ip_{2i-1}T_i/2,3iT_i)\nonumber\\
&+&\theta_2(ip_{2i-1}T_i/2,iT_i)\theta_2(ip_{2i-1}T_i/2,3iT_i)\Bigr] 
\nonumber \\
\label{yuka2}
\end{eqnarray}
\begin{eqnarray}
h_i&\sim&
e^{-\pi(q_i^2+3t_i^2)T_i/12}\Bigl[\theta_3(it_iT_i,4iT_i)\theta_3(iq_iT_i,12
iT_i)\nonumber\\
&+&\theta_2(it_iT_i,4iT_i)\theta_2(iq_iT_i,12iT_i)\Bigr] \nonumber \\
\label{yuka3}
\end{eqnarray}
and
\begin{eqnarray}
h_i&\sim&
e^{-\pi(\tilde{q}_i^2+\tilde{t}_i^2)T_i/12}\Bigl[\theta_3(i\tilde{q}_iT_i,4i
T_i)\theta_3(-i\tilde{t}_iT_i,4iT_i)\nonumber\\
&+&\theta_2(i\tilde{q}_iT_i,4iT_i)\theta_2(-i\tilde{t}_iT_i,4iT_i)\Bigr] 
\nonumber \\
\label{yuka4}
\end{eqnarray}
The Yukawas $h=\prod _{i=1}^{3}h_i$ correspond to the superpotential terms.
 However, the normalisation factors deriving from the K${\rm {\ddot
a}}$hler potential are 
functions of $T_i+\bar{T}_i$ and are not relevant for the present
discussion, because they do not contribute to any phases.
As discussed above, it is of interest to consider the case
$T_1=T_2=T_3=T=e^{i\pi/6}$ 
corresponding to one of the fixed points of the $PSL(2,Z)$ modular group.
Then,the possible 
phases of $h_i$ in (\ref{yuka1})\ldots(\ref{yuka4}) for the allowed values 
of the integers $p_{2i-1},q_i,t_i,\tilde{q}_i$ and $\tilde{t}_i$ are
given in tables 1-4.
For a particular orbifold, the possible phases of $h$ are obtained by
combining the phases 
of $h_i, i=1,2,3$. 
The values obtained are 
of the right 
order of magnitude to yield phases of order $10^{-1}-10^0$ in the 
CKM matrix.

When, as discussed earlier, the non-perturbative superpotential $W_{np}$ is
generalised to 
involve the absolute modular invariant $j(T)$ as well as the Dedekind eta
function $\eta(T)$,
 minimisation of the effective potential can lead to values of $T$ on the
unit circle that 
differ from $1$ or $e^{i\pi/6}$. Then, $W_{np}$ contains an extra factor
$H(T)$, where the 
most general form of $H(T)$ to avoid singularities in the fundamental
domain \cite{CFILQ} is
\begin{equation}
H(T)=(j-1728)^{m/2}j^{n/3}P(j)
\end{equation}
where $m$ and $n$ are positive integers, and $P(j)$ is a polynomial in $j$. 
This
results
 in modification of (\ref{aterm})\ldots(\ref{scalar}) by the replacement of 
$(\tilde{\delta}_{GS}\rho-3)\hat{G}(T,\bar{T})$ by
$(\tilde{\delta}_{GS}\rho-3)\hat{G}(T,\bar{T})
+\frac{dlogH}{dT}$. At least for the case where stabilisation of the
dilaton expectation value
is due to a stringy non-perturbative dilaton  K${\rm {\ddot a}}$hler
potential, there is 
a region of parameter space for which the value of $T$ at the minimum of the
effective potential
 is such that
$(\tilde{\delta}_{GS}\rho-3)\hat{G}(T,\bar{T})+\frac{dlogH}{dT}$ is zero.
Then, 
the $CP$-violating phases in the supersymmetry-breaking terms 
are again zero and the soft supersymmetry-breaking terms remain
universal. However,
 the minimum is no longer at a zero of $\hat{G}(T,\bar{T})$, and
consequently no longer at a 
fixed point of the modular group. We therefore have to evaluate the phases
of the Yukawas at
 a complex value of $T$ different from $e^{i\pi/6}$. For example, for
$P(J)=1, m=n=1,
 \tilde{\delta}_{GS}=-30 , \frac{dP}{dy}=2.45 ,\frac{d^2P}{dy^2}=0.45 $, there is a
minimum with $T$ 
on the unit circle at 
\begin{equation}
T=0.971143367\pm 0.238496458\;i
\label{Tcirc}
\end{equation}  
The range of the phases of $h_i$ are listed in tables 5-8.
Again the phases are 
predominantly 
of order $10^{-1}-10^{0}$.
Of course there are also minima with $T$ obtained by performing a modular
transformation
 on $e^{\frac{i\pi}{6}}$ or 
(\ref{Tcirc}). We have checked that the order of magnitude of the 
largest phase for each variety of Yukawa does not change when 
modular transformations are carried out.
In a full
calculation
 of the CKM matrix from a detailed 3 generation orbifold model, the phases of
 more than one Yukawa coupling will enter and the result for the observable
$CP$-violating
 phases must be modular invariant.

In conclusion, we have studied orbifold compactifications where the value of
the orbifold
 modulus is at a fixed point $e^{i\pi/6}$ of the modular group \cite{FMTV,
Achar, BKL1,BKL2}.
(This happens for quite a wide range of choices of the dilaton auxiliary
field $F_S$ or the 
derivatives $\frac{dP}{dy}$ and $\frac{d^2P}{dy^2}$ of the dilaton 
K${\rm {\ddot a}}$hler potential.) The $CP$-violating phases in the soft 
supersymmetry-breaking terms are zero and the soft supersymmetry-breaking
terms are 
universal \cite{Achar,BKL1,BKL2}. As a result, the contribution to the
electric dipole
 moment of the neutron from the phase of the $T$ modulus is zero. Moreover, 
large contributions to the $\epsilon_K$ parameter in the neutral kaon
system arising 
from the squark mass matrix are avoided
\cite{Gross}. On the other hand, the largest
phase in the CKM 
matrix is expected to be of order $10^{-1} - 10^0$ generically. The picture
of $CP$-violation
 in the neutral kaon system is thus essentially the same as in the standard 
model \cite{Gross} requiring small weak mixing angles to explain the
smallness of the
$CP$-violation in this system and implying larger $CP$-violation in neutral
$B$ decays.
However, unlike the generic situation in low energy supergravity, the
smallness of the
neutron electric dipole moment is rather natural. Similar conclusions can
be drawn when
 the non-perturbative superpotential is generalised to include the absolute
modular
 invariant $j(T)$ when T has complex values on the unit circle that result
in the 
generalisation of $\hat{G}(T,\bar{T})$ to this case being zero.

{\bf Acknowledgements}
This research was supported in part by PPARC.

\newpage

\begin{table}
\begin{center}
\begin{tabular}{|c|c|} \hline\hline
$p_{2i-1}$ & ${\rm Arg}(h_i)$ \\ \hline\hline
0 & 0 \\
$\pm 1$ & $-\pi/3$ \\
$\pm 2$ & $-\pi/3$ \\
\hline\hline
\end{tabular}
\end{center}
\caption{Phases of Yukawa coupling Eq.33  at the fixed point 
$e^{{i\pi}{6}}$ of the modular group}
\end{table}

\begin{table}
\begin{center}
\begin{tabular}{|c|c|} \hline\hline
$p_{2i-1}$ & ${\rm Arg}(h_i)$ \\ \hline\hline
0 & -0.3746 \\
$\pm 1$ &-0.5899 \\
\hline\hline
\end{tabular}
\end{center}
\caption{Phases of Yukawa coupling Eq.34  at the fixed point 
$e^{{i\pi}{6}}$ of the modular group}
\end{table}

\begin{table}
\begin{center}
\begin{tabular}{|c|c|c|c|} \hline\hline
$n_{2i-1}$ & $p_{2i}$ & $p_{2i-1}$ &${\rm Arg}(h_i)$ \\ \hline\hline
-1 & 0 & 0 & $\pi/3$ \\
-1 & 1 & 0 & -0.392  \\
 1 & 0 & 0 & $\pi/3$ \\
 1 & 1 & 1 & -0.982 \\
 0 & 1 & 0 & -0.982  \\
 1 & 1 & 0 & -0.982 \\
 0 & 1 & 1 & -0.982  \\
-1 & 0 & 1 & $-\pi/6$  \\
-1 & 1 & 1 & -0.392 \\
 0 & 0 & 0 & 0.0 \\
 1 & 0 & 1 & $-\pi/6$ \\
 0 & 0 & 1 & $-\pi/2$ \\
\hline\hline
\end{tabular}
\end{center}
\caption{Phases of  Yukawa coupling Eq.35  at the fixed point 
$e^{{i\pi}{6}}$ of the modular group}
\end{table}

\begin{table}
\begin{center}
\begin{tabular}{|c|c|c|c|} \hline
$\tilde{n}_{2i}$ & $\tilde{p}_{2i-1}$ 
& $\tilde{p}_{2i}$ & ${\rm Arg}(h_i)$ \\ \hline\hline
1 & 0 & 0 & $-\pi/2$ \\
1 & 1 & 1 & $-\pi/4$ \\
0 & 0 & 0 & 0 \\
1 & 0 & 1 & $-\pi/4$ \\
0 & 1 & 1 & $-\pi/4$  \\
1 & 1 & 0 & 0 \\
0 & 1 & 0 & $-\pi/2$\\
0 & 0 & 1 & $-\pi/4$\\
\hline\hline
\end{tabular}
\end{center}
\caption{Phases of Yukawa coupling Eq.36  at the fixed point 
$e^{{i\pi}{6}}$ of the modular group}
\end{table}
\newpage
\begin{table}
\begin{center}
\begin{tabular}{|c|c|} \hline\hline
$p_{2i-1}$ & ${\rm Arg}(h_i)$ \\ \hline\hline
0 & -0.013383 \\
$\pm 1$ & -0.50174 \\
$\pm 2$ & -0.50174 \\
\hline\hline
\end{tabular}
\end{center}
\caption{Phases of Yukawa coupling Eq.33 
for minimum of $T$ on the 
unit circle (Eq.38) }
\end{table}

\begin{table}
\begin{center}
\begin{tabular}{|c|c|} \hline\hline
$p_{2i-1}$ & ${\rm Arg}(h_i)$ \\ \hline\hline
0 & -0.159 \\
$\pm 1$ & -0.285 \\
\hline\hline
\end{tabular}
\end{center}
\caption{Phases of Yukawa coupling Eq.34 
for minimum of $T$ on the 
unit circle (Eq.38) }
\end{table}

\begin{table}
\begin{center}
\begin{tabular}{|c|c|c|c|} \hline\hline
$n_{2i-1}$ & $p_{2i}$ & $p_{2i-1}$ &${\rm Arg}(h_i)$ \\ \hline\hline
-1 & 0 & 0 & -0.999 \\
-1 & 1 & 0 & -0.1897  \\
 1 & 0 & 0 & -0.999 \\
 1 & 1 & 1 & -0.4704 \\
 0 & 1 & 0 & -0.4704 \\
 1 & 1 & 0 & -0.4704 \\
 0 & 1 & 1 & -0.4704 \\
-1 & 0 & 1 & -0.25422 \\
-1 & 1 & 1 & -0.1897 \\
 0 & 0 & 0 & 0.0 \\
 1 & 0 & 1 & -0.25422 \\
 0 & 0 & 1 & -0.75149 \\
\hline\hline
\end{tabular}
\end{center}
\caption{Phases of Yukawa coupling Eq.35 for minimum of $T$ on the 
unit circle (Eq.38)}
\end{table}

\begin{table}
\begin{center}
\begin{tabular}{|c|c|c|c|} \hline
$\tilde{n}_{2i}$ & $\tilde{p}_{2i-1}$ 
& $\tilde{p}_{2i}$ & ${\rm Arg}(h_i)$ \\ \hline\hline
1 & 0 & 0 & -0.74926 \\
1 & 1 & 1 & -0.37909 \\
0 & 0 & 0 & -0.0089 \\
1 & 0 & 1 & -0.37909 \\
0 & 1 & 1 & -0.37909\\
1 & 1 & 0 & -0.00892 \\
0 & 1 & 0 & -0.74926\\
0 & 0 & 1 & -0.37909 \\
\hline\hline
\end{tabular}
\end{center}
\caption{Phases of Yukawa coupling Eq.36  for minimum of $T$ on the 
unit circle (Eq.38)}
\end{table}


\begin{thebibliography}{99}
\bibitem{Dine}M. Dine, R.G. Leigh and D.A. MacIntire, Phys. Rev. Lett.
69(1992)2030
\bibitem{Iblust}L.E. Ib${\rm \acute{a}\tilde{n}}$ez, D. L${\rm{\ddot u}}$st, Phys. Lett. B267(1991)51
\bibitem{Brig}A. Brignole, L.E.  Ib${\rm \acute{a}\tilde{n}}$ez 
and C. Mu${\rm \tilde{n}}$oz, Nucl. Phys. B422(1994)125
\bibitem{Achar}B. Acharya, D. Bailin, A. Love, W.A. Sabra and S. Thomas, Phys.
Lett. B357(1995)387
erratum {\it ibid} B407(1997)451
\bibitem{BKL1}D. Bailin, G.V. Kraniotis and A. Love, Phys. Lett. B414(1997)269
\bibitem{BKL2}D. Bailin, G.V. Kraniotis and A. Love, Nucl. Phys. B518(1998)92
\bibitem{Gross}Y. Grossman, Y. Nir and R. Rattazzi, hep-th/9701231
\bibitem{FMTV}S. Ferrara, M. Magnoli, T.R. Taylor and G. Veneziano, Phys. Lett.
B245(1990)409
\bibitem{Casas}J.A. Casas, preprint SCIPP-96-20, IEM-FT-129/96
\bibitem{iblust92}L.E. Ib${\rm \acute{a}\tilde{n}}$ez 
and D. L${\rm{\ddot u}}$st, Nucl. Phys.
B382(1992)305
\bibitem{DKL}L.J. Dixon, V.S. Kaplunovsky and J. Louis, Nucl. Phys. B355(1991)649
\bibitem{DFKZ} J.P. Derendinger, S. Ferrara, C. Kounnas and F. Zwirner, Nucl.
Phys. B372(1992)145
\bibitem{AGNT}I. Antoniadis, E. Gava, K.S. Narain and T.R. Taylor, Nucl. Phys.
B432(1994)187
\bibitem{CM}J. Casas and C. Mu${\rm \tilde{n}}$oz, Nucl. Phys. B332(1990)189, erratum {\it
ibid} B340(1990)280
\bibitem{BLS}D. Bailin, A. Love and W.A. Sabra, Nucl. Phys. B403(1993)265
\bibitem{CFILQ}M. Cveti${\rm \check{c}}$, A. Font, L.E. 
Ib${\rm \acute{a}\tilde{n}}$ez, D. L${\rm{\ddot u}}$st and
F. Quevedo, 
Nucl. Phys. B361(1991)194
\end{thebibliography}
\end{document}